# Characterization of E'$_\delta$ and triplet point defects in oxygen deficient amorphous silicon dioxide

G. Buscarino,*  S. Agnello and  F. M. Gelardi

Department of Physical and Astronomical Sciences, University of Palermo, Via Archirafi 36, I-90123 Palermo, Italy

## ABSTRACT

We report an experimental study by electron paramagnetic resonance (EPR) of γ ray irradiation induced point defects in oxygen deficient amorphous $SiO_2$ materials. We have found that three intrinsic (E'$_\gamma$, E'$_\delta$ and triplet) and one extrinsic ($[AlO_4]^0$) paramagnetic centers are induced. All the paramagnetic defects but E'$_\gamma$ center are found to reach a concentration limit value for doses above $10^3$ kGy, suggesting a generation process from precursors. Isochronal thermal treatments of a sample irradiated at $10^3$ kGy have shown that for T≥500 K the concentrations of E'$_\gamma$ and E'$_\delta$ centers increase concomitantly to the decrease of $[AlO_4]^0$. This occurrence speaks for an hole transfer process from $[AlO_4]^0$ centers to diamagnetic precursors of E' centers proving the positive charge state of the thermally induced E'$_\gamma$ and E'$_\delta$ centers and giving insight on the origin of E'$_\gamma$ from an oxygen vacancy. A comparative study of the E'$_\delta$ center and of the 10 mT doublet EPR signals on three distinct materials subjected to isochronal and isothermal treatments, has shown a quite general linear correlation between these two EPR signals. This result confirms the attribution of the 10 mT doublet to the hyperfine structure of the E'$_\delta$ center, originating from the interaction of the unpaired electron with a nucleus of $^{29}Si$ (I=1/2). Analogies between the microwave saturation properties of E'$_\gamma$ and E'$_\delta$ centers and between those of their hyperfine structures are found and suggest that the unpaired electron wave function involves similar Si $sp^3$ hybrid orbitals; specifically, for the E'$_\delta$ the unpaired electron is supposed to be delocalized over four such orbitals of four equivalent Si atoms. Information on the structural model of the triplet center are also obtained indicating that it could consists of the same microscopic structure as the E'$_\delta$ but for a doubly ionized state.

# I. INTRODUCTION

Amorphous silicon dioxide (a-SiO$_2$) is a widely employed material in many optical and electronic applications especially in modern MOS (Metal-Oxide-Semiconductor) devices.[1-3] A drawback to the wide use of the material is that its exposure to ionizing radiation induces the appearance of point defects determining the degradation of the material and of the proper work of the devices.[1,3-4] Among the induced defects, the paramagnetic E' centers play a crucial role since they act as charge traps and can cause optical absorption.[1,2] The most relevant of these defects are the E'$_\gamma$ and E'$_\delta$ centers, whose microscopic structures are still questioned.

The E'$_\gamma$ center is characterized by an almost axially symmetric EPR lineshape around g~2.0006 and a correlated doublet of EPR lines split by ~42 mT, arising from the hyperfine interaction of the unpaired electron with a $^{29}$Si nucleus (4.7% natural abundant isotope with nuclear spin I=1/2).[5] These EPR features of the defect are explained assuming that the unpaired electron is localized in a sp$^3$ hybrid orbital of a threefold coordinated silicon atom.[5-7] The microscopic structure of the defect has been widely studied and its most accepted model consists in a positively charged puckered oxygen vacancy: O≡Si$^\bullet$ $^+$Si≡O (where ≡ represents the bonds to three oxygen atoms, $^\bullet$ represents an unpaired electron and $^+$ is a trapped hole.)[1,2,5,7] On the basis of this microscopic model, the E'$_\gamma$ center is considered as the equivalent in a-SiO$_2$ of the E'$_1$ center of quartz.[5-11] However, a definitive attribution of the microscopic model for the E'$_\gamma$ is still lacking since a correlation with the precursor defect, the oxygen vacancy, has not yet been clearly posed and other precursors have also been suggested.[1, 2]

The E'$_\delta$ center has been observed in bulk a-SiO$_2$[12-14] and in a-SiO$_2$ layers[15-21] on silicon. Its EPR resonance line is nearly isotropic (g~2.002) and together with this signal a pair of lines split by ~10 mT, supposed to arise from hyperfine interaction of the unpaired electron with a $^{29}$Si nucleus, is usually detected.[12,14] The E'$_\delta$ center has been verified to have an hole-trap nature in SiO$_2$ films on silicon by capacitance-voltage measurements and charge injection.[16,18,19] Until now four distinct models have been proposed for the E'$_\delta$ and their main features are here summarized. The model proposed by Griscom and Friebele consists in an electron delocalized over an [SiO$_4$]$^{4+}$ vacancy decorated by three Cl$^-$ ions (4-Si Cl-containing model).[12] Tohmon *et al.*[13] have pointed out that oxygen deficiency of the material is necessary for E'$_\delta$ formation, and have proposed a model consisting in a ionized single oxygen vacancy with the unpaired electron nearly equally shared by the two Si atoms (2-Si model). Vanheusden and Stesmans[15] have suggested that E'$_\delta$ could consists in a ionized silicon nanocluster of five Si atoms (5-Si cluster model) basing on the observation that this defect occurs nearer to the interface with the Si substrate in SIMOX materials with respect to the E'$_\gamma$ center. Zhang and Leisure,[14] on the basis of the intensity of the 10 mT hyperfine doublet, estimated by high-power second-harmonic EPR measurements, have proposed a model for E'$_\delta$ center consisting in a [SiO$_4$]$^+$ vacancy (4-Si model) with the unpaired electron delocalized over the four nearby Si atoms.

Another characteristic EPR signal with g~4 is also found in the same materials in which the E'$_\delta$ center is revealed.[12-14] This resonance has been attributed to a weakly allowed transition between the states |m$_s$=-1> and |m$_s$=+1> of a coupled spins system with total spin S=1 and $\hat{S}_z$ eigenvalues m$_s$=-1, 0, +1.[12] For pure |m$_s$=-1> and |m$_s$=+1> states and for a microwave magnetic field orthogonal to the static magnetic field, no first order transition can occur between these two states.[22] In a conventional EPR spectrometer the latter condition is always satisfied. However, for large values of the *zero field splitting*, D,[22] the three spin states are linear combinations of the pure |m$_s$=-1>, |m$_s$=0>, |m$_s$=+1> states, consequently the transition between the lowest and the highest energy levels can be observed and gives rise to an EPR line at g~4.[22] Griscom and Friebele[12] have isolated also a pair of lines with peak-to-peak splitting of ~13 mT, attributed to the allowed transitions between the states |m$_s$=-1> ↔ |m$_s$=0> and |m$_s$=0> ↔ |m$_s$=+1> of the triplet center, and have

estimated a *zero field splitting* D≅13.4 mT. This large value of D is actually responsible for the occurrence of the resonance at g~4.

As a consequence of the observation of a similar growth of concentration with increasing X-ray irradiation dose it has been supposed that the triplet center could share the same precursor of E'$_\delta$ center.[14] In this scheme, a single and a double ionization of the same precursor site could originate the E'$_\delta$ and the triplet center, respectively.[12-14]

In the present paper, which expands the topic of a recent Letter,[23] we extend our study describing the growth of concentration of E'$_\gamma$, E'$_\delta$, and triplet centers with γ-ray irradiation dose. To better characterize all these defects the room temperature saturation properties with microwave power of their first-harmonic EPR signals are also studied. The latter studies give indication on the efficiency of the relaxation processes of a given paramagnetic center, that is a key property for the identification and the characterization of a point defect, and have enabled us to find new evidences on the microscopic structure of the E'$_\delta$ center. We report isochronal annealing experiments that have pointed out that an hole transfer process from [AlO$_4$]$^0$ centers to the sites precursor of E' occurs, generating E'$_\gamma$ and E'$_\delta$ centers and proving the positive charge state of these induced defects. Furthermore, we have found that EPR signals of the E'$_\delta$ main resonance and of the 10 mT doublet change in a strongly correlated way during thermal treatments, confirming the attribution of the latter to the hyperfine structure of the E'$_\delta$ center.

## II. EXPERIMENTAL DETAILS

All the materials considered here are commercial a-SiO$_2$. Two of these are obtained from fused quartz, QC and Pursil 453,[24] while a third material, KUVI,[25] is synthesized by vapour axial deposition technique. The optical absorption spectra of these materials show an intense band peaked at ~7.6 eV of amplitude of ~20 cm$^{-1}$ for KUVI, and larger than 100 cm$^{-1}$ for Pursil 453 and QC, characterizing them as oxygen deficient silicon dioxide.[26] Furthermore, all these materials have an Al atoms content of about 10$^{17}$ cm$^{-3}$.[24,25] γ ray irradiation has been carried out at room temperature and with dose rate ~7 kGy/hr. Different samples of Pursil 453 were irradiated in the dose range from 5 kGy to 10$^4$ kGy. Successively, a sample of this materials irradiated at a dose of ~10$^3$ kGy was subjected to an isochronal thermal treatment from 330 K to 800 K (the details of this experiment are reported in ref. 23). Two samples of the same KUVI material, hereafter referred to as KUVI/1 and KUVI/2, were simultaneously irradiated at a dose of ~124 kGy and were successively subjected to a series of isothermal treatments at fixed temperatures of 580 K and 630 K, respectively. A sample of QC, irradiated at ~73 kGy, was isothermally treated at 630 K. In all the isothermal treatment experiments the sample was kept at a fixed temperature for a time $t_0$ and then was cooled to room temperature to perform the EPR measurements. The time $t_0$ was varied from 30 seconds up to many minutes with a sequence depending on the experiment. Finally, to study the microwave saturation properties of the hyperfine doublet of the E'$_\gamma$ center split by ~42 mT, a sample of fused quartz EQ906,[24] γ ray irradiated at a dose of ~10$^3$ kGy, was also considered. In this material no absorption band peaked at ~7.6 eV was detected. All the samples considered in the present work have size 5x5x1 mm$^3$.

EPR measurements were carried out at room temperature with a Bruker EMX spectrometer working at frequency ν ≈ 9.8 GHz (X-band) and with magnetic-field modulation frequency of 100 kHz detecting the first-harmonic unsaturated mode (FH-EPR) or the high-power second-harmonic mode (SH-EPR). The latter measurements were used to reveal the 10 mT hyperfine doublet when a large sensitivity was required. Concentration of defects was determined, with relative accuracy of 10%, by double integration of the FH-EPR spectra and by comparison with the double integral of E'$_\gamma$ center in a reference sample. The defects concentration in the latter was evaluated, with absolute accuracy of 20%, using the instantaneous diffusion method in spin-echo decay measurements carried out in a pulsed EPR spectrometer.[27]

III. EXPERIMENTAL RESULTS

A. EFFECTS OF γ RAY IRRADIATION

1. E'$_\gamma$ and E'$_\delta$ centers

In all the samples no EPR signal was detected before irradiation. At variance, after irradiation EPR signals are induced, as shows the spectrum of Fig. 1(a) centered in correspondence to g~2 and obtained for a sample of Pursil 453 irradiated at $10^3$ kGy (continuous line). This EPR signal arises from the partial superposition of two distinct resonance lines ascribed to E'$_\gamma$ and E'$_\delta$ centers.[12] We have separated these two contributions by fitting the spectrum with a weighted sum of an experimental line shape for E'$_\gamma$ center [Fig. 1(c)],[28] and a simulated line shape for E'$_\delta$ center [Fig. 1(b)]. The latter was obtained by the Bruker's SimFonia software. The result of this procedure is reported in Fig. 1(a), where the weighted sum (circles) of the reference line shapes for E'$_\gamma$ and E'$_\delta$ is superimposed to the experimental spectrum (continuous line). From the analysis reported in Fig. 1, and fixing $g_{||}$=2.0018 for E'$_\gamma$,[1] a zero crossing g value of 2.0020±0.0001 has been obtained for E'$_\delta$ center, in good agreement with other experimental estimations.[12,15,16,18-20]

The line shapes reported in Fig. 1(b) and 1(c) were also used to estimate the concentrations of E'$_\gamma$ and E'$_\delta$ induced in other samples of Pursil 453 irradiated to different gamma ray doses. The results obtained are reported in Fig. 2. The concentration of E'$_\delta$ centers was found to increase with irradiation dose up to ~$10^2$ kGy. For higher doses a maximum concentration of ~$10^{16}$ spins/cm$^3$ is maintained, suggesting a generation process from precursor defects.

The concentration of E'$_\gamma$ centers increases up to the highest dose considered, indicating a more complex generation process that could involve a direct activation of normal matrix sites or a not complete exhaustion of precursor defects. Similar concentration growths were previously reported for an X-ray irradiated synthetic a-SiO$_2$ material.[14]

To better characterize E'$_\gamma$ and E'$_\delta$ defects, the room temperature saturation properties of their FH-EPR signals with microwave power were studied. These data are reported in Fig. 3 for the sample of Pursil 453 irradiated at 10$^4$ kGy and point out that the two E' centers have virtually identical saturation properties. Moreover, these saturation curves also reproduce those reported for Type I-IV commercial a-SiO$_2$[12,29].

The hyperfine doublets split by 42 mT and 10 mT, associated to E'$_\gamma$ and E'$_\delta$ centers respectively, were also investigated by EPR measurements in the samples of Pursil 453. However, due to the low concentration of defects, a quantitative analysis was prevented.

**2. Triplet center**

In the irradiated samples of Pursil 453 we looked for the g~4 resonance of the triplet center. To this aim we have performed measurements setting the magnetic field at approximately half of the resonance field of the E' centers. As reported in Fig. 4 for the sample irradiated at 10$^4$ kGy, a FH-EPR signal was detected with line shape and resonace magnetic field compatible with those ascribed to the triplet center.[12,13]

To obtain an estimation of the triplet centers concentration, the intensity of the FH-EPR lines split by ~13 mT due to the allowed transitions between the states $|m_s=-1\rangle \leftrightarrow |m_s=0\rangle$ and $|m_s=0\rangle \leftrightarrow |m_s=+1\rangle$ had to be determined. In our samples, due to the presence of the intense EPR signal of the $[AlO_4]^0$ centers (vide infra), we were not able to isolate these lines. However, since it was reported for the triplet center in a-SiO$_2$ that the ~13 mT pair is ~2500 times more intense than the g~4 resonance,[12] we have estimated the concentration of triplet centers multiplying by a factor 2500 the double integral of the g~4 FH-EPR signal. The value obtained for various irradiation doses are reported in Fig. 2. From the comparison of the growth characteristics of E'$_\delta$ and triplet centers, it is evident that the maximum value of concentration is reached at the same dose value. The concomitant presence of E'$_\delta$ and triplet centers in our material, together with the analogy in the growth of concentration with irradiation dose, indicates the existence of some correlation between these two centers. In particular, these features are compatible with the hypothesis that these two centers could share the same precursor.[12-14]

The dependence of the FH-EPR signal on the microwave power for the g~4 resonance is reported in Fig. 5. The EPR signal was found to grow *linearly* with the microwave power up to ~50 mW. Deviation from the linear dependence was observed for higher power due to the occurrence of saturation effects. We note that the saturation occurs at higher power with respect to E'$_\gamma$ and E'$_\delta$, indicating that the triplet center possesses more effective relaxation channels with respect to the E' centers.

## 3. $[AlO_4]^0$ center

After γ-ray irradiation another paramagnetic center was also induced in our samples of Pursil 453. The value of the resonant magnetic field and the characteristics of the FH-EPR line shape, have permitted us to associate this resonance to the $[AlO_4]^0$ center.[30,31] Experimental[32-34] and theoretical[35,36] studies in quartz have shown that this defect consists in a Al atom substituting for a four-coordinated Si atom in the lattice with a hole trapped in a nonbonding 2p orbital of an O atom adjacent to Al. The existence of the analogous defect in a-SiO$_2$ was also verified.[30,31] In Fig. 2 the growth of concentration of this impurity center on increasing irradiation dose is reported. As shown in the figure, the $[AlO_4]^0$ center concentration was found to increase up to ~$10^3$ kGy, for higher doses a limit value of ~$2 \times 10^{17}$ spins/cm$^3$ is maintained. Since this concentration of defects is comparable with the nominal Al content of the material, we conclude that almost all of the Al atoms present in the material are substitutional of Si and, after irradiation, give rise to the $[AlO_4]^0$ center.

## B. THERMAL TREATMENTS EXPERIMENTS

### 1. Main thermal treatments effects

A sample of Pursil 453 irradiated at $10^3$ kGy was subjected to isochronal thermal treatments and the concentrations of defects as a function of the treatment temperatures are reported in Fig. 6. These data show that $E'_\gamma$, $E'_\delta$ and triplet centers start to anneal at T~400 K. However, while at higher temperature the triplet center anneals out definitively, the $E'_\gamma$ and $E'_\delta$ centers concentrations begin to increase for T≅500 K, indicating that a production mechanism is activated. Maximum concentrations ~3÷4 times larger than the initial values are reached after treatments at T~580 K and T~620 K for $E'_\delta$ and $E'_\gamma$, respectively. For higher temperature first $E'_\delta$, at T~590 K, and then $E'_\gamma$, at T~660 K, anneal. We note that the rate of annealing at high temperature are very different for $E'_\gamma$ and $E'_\delta$ centers, indicating that different processes are involved for the two defects.

Quite different annealing features were found for $[AlO_4]^0$ centers. Thermal treatments up to T~500 K do not significantly change the concentration of defects, while for higher temperature the number of defects decreases. As shown in Fig. 6, this impurity center undergoes a more rapid annealing with respect to that of E' centers. Furthermore, $[AlO_4]^0$ centers anneal out in the same temperature range in which the growth of $E'_\delta$ and $E'_\gamma$ centers occurs and, after each thermal treatment, the total number of the generated E' centers is less than that of annealed $[AlO_4]^0$ centers.

As reported in Ref. 23, the 10 mT doublet signal increases for thermal treatments in the range of temperatures from 450 K to 580 K. Taking advantage of the high signal to noise ratio reached during the thermal treatment experiment, the dependence of the FH-EPR intensity of the 10 mT doublet on microwave power has been investigated. As reported in Fig. 7 for the sample treated at 580 K, where the maximum signal amplitude is recorded, the FH-EPR signal shows a linear growth with increasing microwave power up to ~ $2.5 \times 10^{-2}$ mW, while for higher power a deviation from linear dependence is observed.

For sake of comparison, the 42 mT doublet, hyperfine structure of E'$_\gamma$ center, has been also investigated in the same material. However, due to the superposition with other EPR signals, a reliable saturation curve with microwave power was not obtained. For this reason, we have considered an irradiated sample of EQ906 in which a concentration of ~$10^{17}$ spins/cm$^3$ of E'$_\gamma$ (and no E'$_\delta$ and triplet centers) have been detected. In this sample the room temperature saturation curve with microwave power of the 42 mT doublet was obtained and is reported in Fig. 7. The comparison reported in this figure point out that not only the saturation properties of E'$_\gamma$ and E'$_\delta$ centers are similar, but also those of their hyperfine structures.

## 2. Correlation between the EPR signals of the E'$_\delta$ center and of the 10 mT doublet

To support the attribution of the 10 mT doublet to the hyperfine structure of the E'$_\delta$ center, we have performed a comparative study of the E'$_\delta$ center and of the 10 mT doublet EPR signals in three distinct materials subjected to isochronal and isothermal treatments at two distinct temperatures.

We have verified that an increase of E'$_\gamma$ and E'$_\delta$ centers' EPR signal similar to the one discussed above for the Pursil 453 occurs also in the samples KUVI/1, KUVI/2 and QC, isothermally treated at 580 K, 630 K and 630 K, respectively. In particular, we have found that the concentration of E'$_\delta$ centers and of the 10 mT pair grow up to a total time of the isothermal treatment of ~500 seconds, after that both signals progressively anneal out. In all the irradiated samples the g~4 resonance of the triplet was also detected. However, we have verified that after a treatment of about 60 seconds above 500 K this EPR signal disappears definitively, ruling out a possible contribution under the 10 mT doublet coming from the $|m_s=-1\rangle \leftrightarrow |m_s=0\rangle$ and $|m_s=0\rangle \leftrightarrow |m_s=+1\rangle$ transitions of the triplet center.[12]

The EPR signals intensities of the E'$_\delta$ and of the 10 mT doublet were estimated with the fit procedures described in Fig. 1 and Fig. 8(a), respectively. The latter figure points out that the right component of the 10 mT doublet can be properly fitted by a superposition of three Gaussian profiles: one describes the tail on the left of the spectrum, while the other two Gaussians take into account the right components of the 7.4 mT and of the 10 mT doublets.[23] The SH-EPR intensity of the right component of the 10 mT doublet was obtained by simple integration of the Gaussian profile peaked at ~354 mT. With a similar procedure the SH-EPR signal intensity of the left component of the 10 mT doublet was also estimated, and the total intensity was obtained by summing the contributions of the two components.

In Fig. 8(b) we report the SH-EPR signal of the 10 mT doublet in the samples Pursil 453, KUVI/1, KUVI/2 and QC as a function of the $E'_\delta$ center main line FH-EPR signal, as estimated during the thermal treatments experiments. In the figure, the two EPR signals show a strict correlation for an overall variation of their amplitudes of more than one order of magnitude. This result strongly supports the attribution of the 10 mT doublet to the hyperfine structure of the $E'_\delta$ center, arising from the hyperfine interaction of the unpaired electron of the $E'_\delta$ center with a nucleus of $^{29}Si$ (I=1/2).

## IV. DISCUSSION

The data reported here on the isochronal thermal treatments evidence that $E'_\delta$ and $E'_\gamma$ are induced at the same time as the $[AlO_4]^0$ centers are annealed. A similar temperature dependence, occurring in the same temperature range, is typically observed in quartz.[10] In that case, by a detailed EPR analysis, an hole transfer process from $[AlO_4]^0$ to the precursors of $E'_1$ center was supposed.[10] The analogies between the annealing features observed in our samples with that reported for quartz, suggest that a similar process could occur in a-$SiO_2$ materials. Furthermore, the generation of E' centers by a hole transfer process indicates that the defects induced in our bulk sample are positively charged, as previously pointed out for film samples.[16,18,19] In this respect the thermally induced $E'_\gamma$ center could be considered the direct analogous in a-SiO$_2$ of the E'$_1$ center of quartz. In more details, the E'$_\gamma$ center induced during annealing should originate from an oxygen vacancy that, by trapping an hole, becomes paramagnetic, as well as the E'$_1$.

The annealing curves of E'$_\delta$ and E'$_\gamma$ centers during the isochronal thermal treatment experiment were found to be quite different. As shown in Fig. 6, in fact, the former increases up to T~580 K whereas the latter up to T~620 K. Furthermore, for higher temperatures the rate of annealing is different for the two defects. These features should be considered an evidence of the different precursors of these defects. In fact, if an oxygen vacancy were a precursor for both E'$_\delta$ and E'$_\gamma$, there should be no reason to observe temperature differences in their increase, since the common generation process of hole trapping. Also, if these two defects consist in a ionized oxygen vacancy then a similar annealing rate is expected, in contrast with our finding. On these bases it can be guessed that the single oxygen vacancy, precursor of the E'$_\gamma$, is not a reliable precursor for the E'$_\delta$ center.

As shown in Fig. 3, E'$_\gamma$ and E'$_\delta$ main resonances feature essentially the same microwave power dependences. Interestingly, as reported in Fig. 7, a similar result has been also obtained for their hyperfine structures. These analogies point out that similar relaxation mechanisms occur for the two defects, and suggest that their unpaired electrons should essentially be located in similar sp$^3$ orbitals.

Our previous work enabled us to discern among the various models of E'$_\delta$,[23] basing on the ratio:

$$\zeta = \frac{\text{hyperfine doublet FH–EPR intensity}}{\text{main resonance FH–EPR intensity}}$$

$$= 0.047 \cdot n \cdot (1 - 0.047)^{(n-1)},$$

where 0.047 is the natural abundance of $^{29}$Si nuclei and n indicates that the unpaired electron wave function is delocalized over n Si atoms. In Ref. 23 we have obtained $\zeta$ =0.16±0.02 which is consistent with the value $\zeta$=0.163 expected for n=4. This outcome indicate that the unpaired electron wave function of the E'$_\delta$ center is actually delocalized over four nearly equivalent silicon atoms. In the model of Vanheusden and Stesmans[15] the unpaired electron was supposed to possess a wave function resulting from the superposition of four $sp^3$ orbitals of the silicon atom disposed at the center of the cluster. However, this model leads to a value of $\zeta \cong 0.047$,[14] in disagreement with our estimation.[23] At variance, if one assumes a complementary view in which the unpaired electron is supposed to be delocalized over the outermost four Si atoms of the 5-Si cluster, the unpaired electron should be visualized at any given time as localized in an $sp^3$ hybrid orbital similar to the one involved in the E'$_\gamma$ center. The overall orbital should consists in a wave function composed by the four $sp^3$ orbitals of the nearby Si atoms. This picture is compatible with the analogy found in the continuous-wave microwave saturation properties of E'$_\delta$ and E'$_\gamma$ reported here and with the expected value of $\zeta$.[23] It is worth to note that the conjecture that E'$_\delta$ and E'$_\gamma$ possess similar $sp^3$ orbitals also agrees with the observed splitting of the hyperfine doublet associated to the E'$_\delta$ center. In fact, under this hypothesis, the hyperfine doublet of the latter is expected to have a splitting of about ¼ · 42 mT $\cong$ 10 mT, due to the electron delocalization over four nearly equivalent Si atoms. The alternative microscopic structure compatible with our experimental findings is the 4-Si model (Zhang and Leisure[14]). Also in this model, in fact, the unpaired electron is supposed to be delocalized over four $sp^3$ orbitals of the nearby Si atoms. Summarizing, our experimental data suggest that the E'$_\delta$ center could originates from a radiation induced ionization of an [SiO$_4$] vacancy [Fig. 9(a)] or of a 5-Si cluster [Fig. 9(d)].[14,15] Irradiation removes an electron from one of the Si-Si bonds of the precursor and after a dynamical relaxation the remaining unpaired electron becomes delocalized over four $sp^3$ hybrid orbitals of the nearby silicon atoms [Fig. 9(b) and Fig. 9(e)].

In the present work we report a new evidence of the concomitant production and of the similar concentration growth for E'$_\delta$ and triplet centers, indicating a correlation between these point defects. If, as already suggested,[12-14] a similar precursor is supposed to be responsible for the generation of E'$_\delta$ and triplet centers, then the latter defect could consists in two weakly interacting unpaired electrons localized in two different Si $sp^3$ orbitals within an [SiO$_4$] vacancy [Fig. 9(c)] or a 5-Si cluster [Fig. 9(f)]. In this scheme, a single and a double ionization of the same precursor site could be the processes responsible for the generation of the E'$_\delta$ and the triplet center, respectively.

## V. CONCLUSIONS

We have reported the growth characteristics of E'$_\gamma$, E'$_\delta$, triplet and [AlO$_4$]$^0$ centers induced by γ ray irradiation of an oxygen deficient a-SiO$_2$ material. Thermal treatments experiments have permitted us to point out that, for temperature above ~500 K, an hole transfer process from [AlO$_4$]$^0$ centers to the sites precursors of E'$_\gamma$ and E'$_\delta$ occurs. The growth of concentration of E'$_\gamma$ and E'$_\delta$ centers induced by hole transfer process gives the first reported evidence of the positive charge state of E'$_\gamma$ and E'$_\delta$ centers in bulk a-SiO$_2$ and suggests for the former a generation from an oxygen vacancy. Furthermore, the variations of the E'$_\delta$ center EPR signal intensity during thermal treatments experiments were found to be strictly correlated to that of the 10 mT doublet, confirming the attribution of the latter to the hyperfine structure of the E'$_\delta$ center.

The study of the microwave saturation properties of E'$_\gamma$ and E'$_\delta$ have pointed out strict analogies both for their main resonance lines and hyperfine structures. This result suggests that the unpaired electron of E'$_\delta$ have similar relaxation properties as the E'$_\gamma$ and that analogous sp$^3$ orbitals are involved in both defects. On the basis of this experimental evidence the unpaired electron wave function of the E'$_\delta$ center is supposed to consist in four sp$^3$ hybrid orbitals of the nearby Si atoms of an [SiO$_4$] vacancy or of a 5-Si cluster. Finally, new evidence has been reported of the correlated generation of the E'$_\delta$ and the triplet center, suggesting that the latter consists of the same microscopic structure as the E'$_\delta$ center but for a doubly ionized state.

# ACKNOWLEDGMENTS

We thank R. Boscaino, M. Cannas, M. Leone, F. Messina for useful discussions and suggestions, E. Calderaro and A. Parlato for taking care of the γ irradiation in the irradiator IGS-3 at the Nuclear Department of Engineering, University of Palermo. This work was financially supported by Italian Ministry of University Research and Technology.

**FIGURE CAPTION**

FIG. 1. (a) FH-EPR spectrum for the sample of Pursil 453 irradiated at $10^3$ kGy acquired in correspondence of g~2 (continuous line) compared to the line obtained as a weighted sum (circles) of the reference lines for E'$_\delta$ (b) and E'$_\gamma$ (c) centers.

FIG. 2. Concentration of γ-ray irradiation induced paramagnetic defects in Pursil 453.

FIG. 3. Room temperature saturation with microwave power of the FH-EPR signal for E'$_\gamma$ (circles) and E'$_\delta$ (stars) centers induced in the sample of Pursil 453 irradiated at $10^4$ kGy. The solid line evidences the linear region of signal growth with microwave power.

FIG. 4. FH-EPR spectrum for the sample of Pursil 453 irradiated at $10^4$ kGy acquired in correspondence of g~4.

FIG. 5. Room temperature saturation with microwave power of the FH-EPR signal for the g~4 resonance of the triplet center induced in the sample of Pursil 453 irradiated at $10^4$ kGy. The solid line evidences the linear region of signal growth with microwave power.

FIG. 6. Concentrations of the paramagnetic defects in the sample of Pursil 453 irradiated at $10^3$ kGy as a function of the temperature of the isochronal thermal treatment.

FIG. 7. Room temperature saturation with microwave power of the FH-EPR signal of the 10 mT doublet in the sample of Pursil 453 irradiated at $10^3$ kGy thermally treated at ~580 K (stars) and of the 42 mT doublet in the sample of EQ906 irradiated at $10^3$ kGy (circles). The solid line evidences the linear region of signal growth with microwave power.

FIG. 8. (a) Right component of the 10 mT doublet detected by SH-EPR measurements for a sample of Pursil 453 irradiated at $\sim 10^3$ kGy and subjected to isochronal thermal treatments up to ~580 K. Broken lines indicate the three Gaussians components used to fit the experimental data. (b) SH-EPR total intensity of the 10 mT doublet as a function of the FH-EPR signal intensity of the E'$_\delta$ center main resonance for the thermally treated samples. The dimension of the symbols are comparable with the error on the measurements. The straight line, with slope 1, is superimposed to the data, for comparison.

FIG. 9. (Color online) 4-Si model for the site precursor of E'$_\delta$ and triplet centers (a), E'$_\delta$ center (b) and triplet center (c). 5-Si model for the site precursor of E'$_\delta$ and triplet centers (d), E'$_\delta$ center (e) and triplet center (f). Arrows represent unpaired electrons in Si sp$^3$ orbitals.

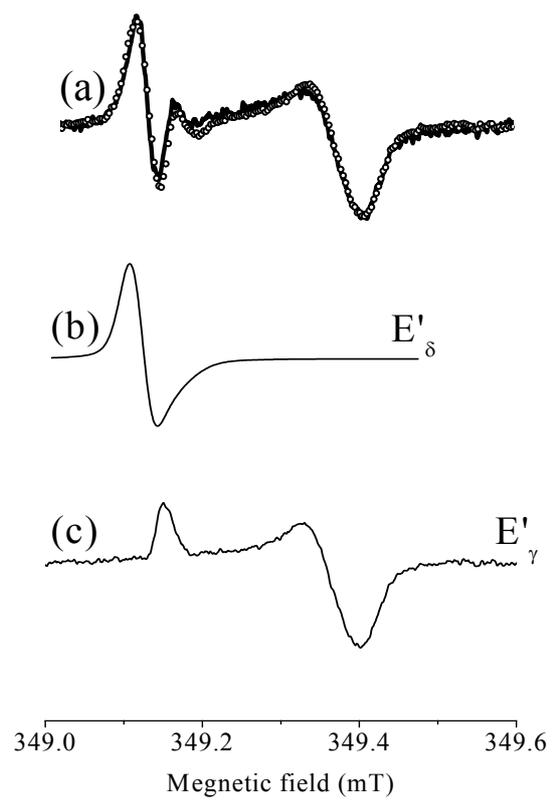

FIGURE 1
BUSCARINO G. et al.

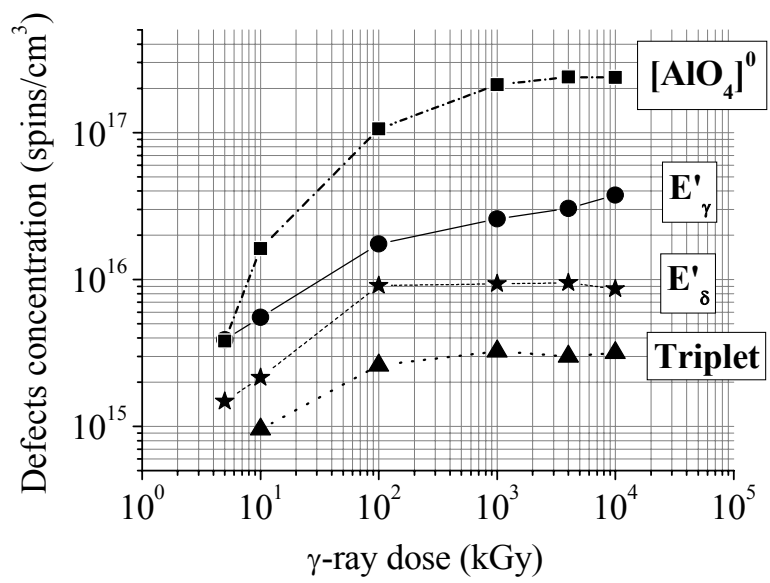

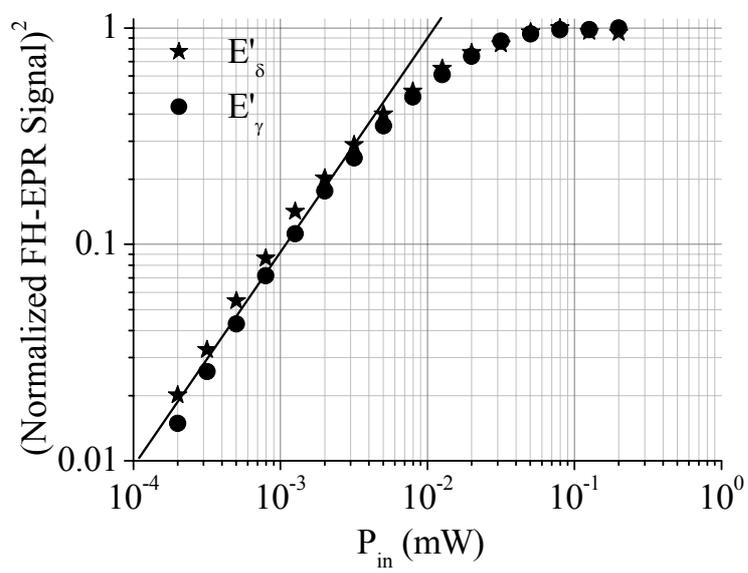

FIGURE 3
BUSCARINO G. et al.

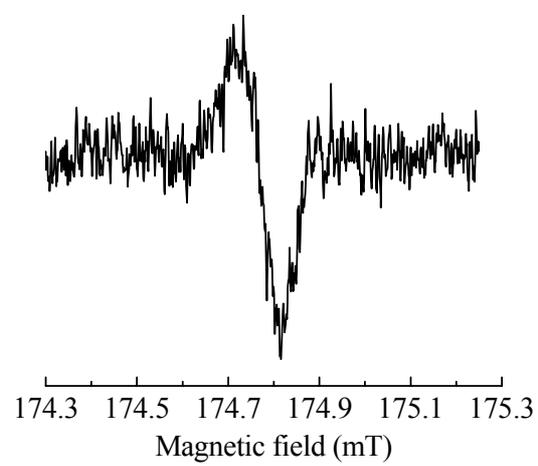

FIGURE 4
BUSCARINO G. et al.

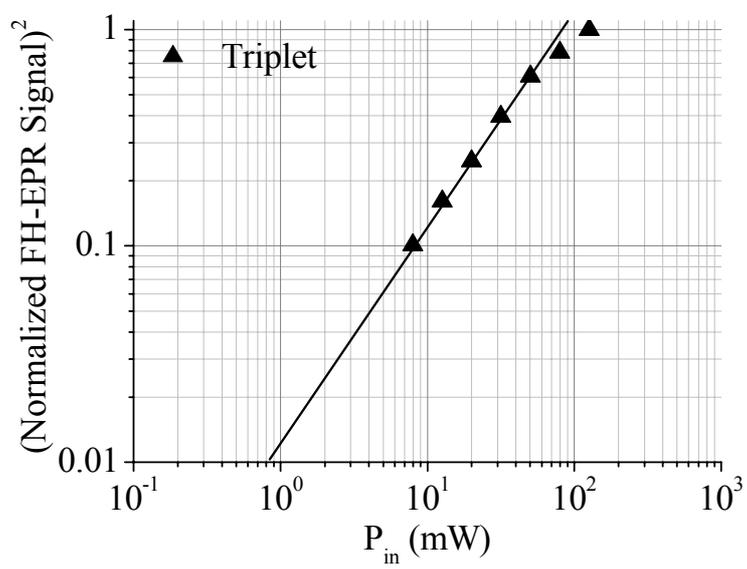

FIGURE 5
BUSCARINO G. et al.

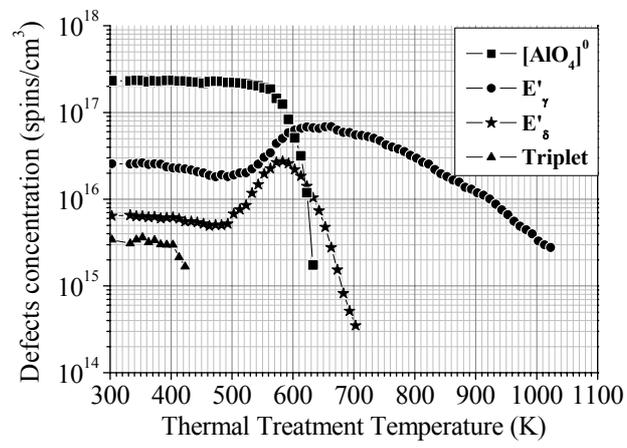

FIGURE 6
BUSCARINO G. et al.

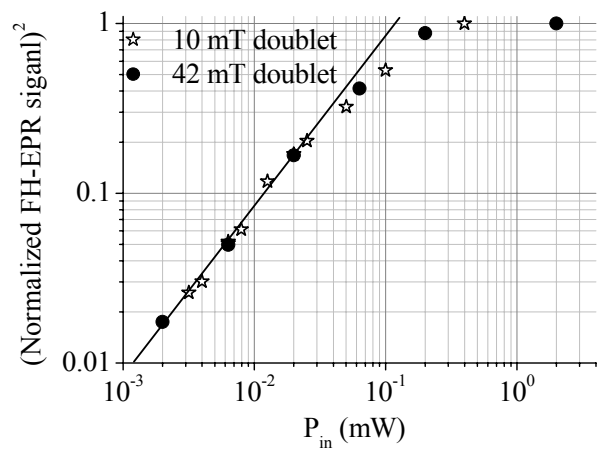

FIGURE 7
BUSCARINO G. et al.

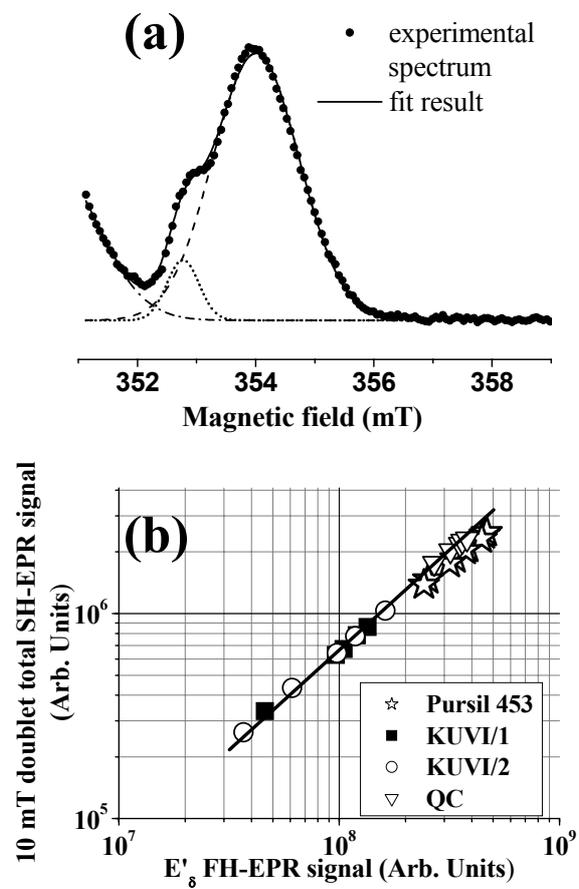

FIGURE 8
BUSCARINO G. et al.

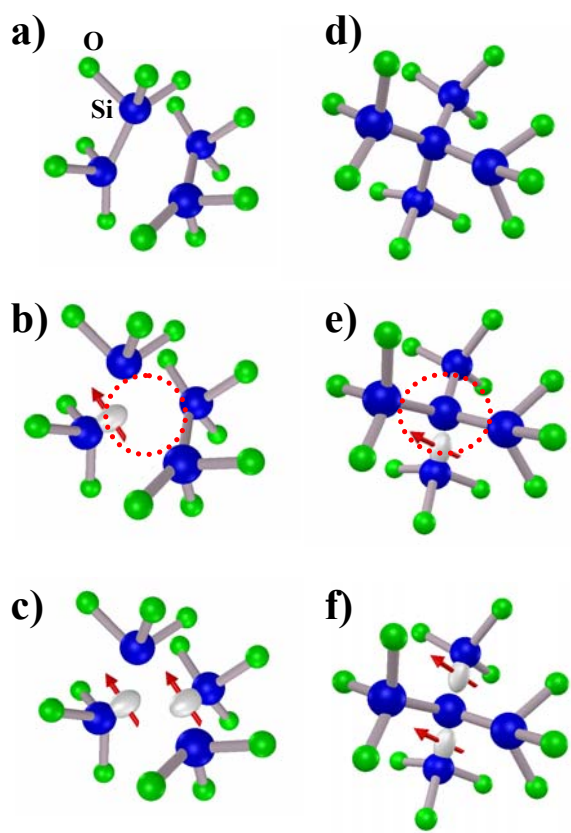

FIGURE 9
BUSCARINO G. et al.

28